\documentclass[lettersize,journal]{IEEEtran}
\usepackage{amsmath,amsfonts}
\usepackage{algorithm}
\usepackage{array}
\usepackage[caption=false,font=normalsize,labelfont=sf,textfont=sf]{subfig}
\usepackage{textcomp}
\usepackage{stfloats}
\usepackage{url}
\usepackage{verbatim}
\usepackage{graphicx}
\usepackage{cite}
\usepackage{algpseudocode}
\usepackage{amsmath}
\usepackage{amssymb}
\usepackage[table]{xcolor}
\usepackage{xcolor}

\usepackage{multirow}
\usepackage{booktabs}

\usepackage{hyperref}
\hypersetup{hidelinks}

\newcommand{\red}[1]{\textcolor{red}{#1}}

\algblockdefx{With}{EndWith}[1]%
  {\textbf{with} #1 \textbf{do}}
  {\textbf{end with}}           

\algrenewcommand\algorithmiccomment[1]{\hfill \textcolor{gray}{\(\triangleright\) \textit{#1}}}

\begin{document}

\title{SwiftAudio: Data-Efficient Caption-Only Distillation for One-Step Text-to-Audio Diffusion-based Generation}

\author{
    Binh Mai,
    Tran Quoc Bao Le,
    Hung Dinh,
    and Cong Tran
\thanks{
Binh Mai, Tran Quoc Bao Le, Hung Dinh, and Cong Tran are with Posts and Telecommunications Institute of Technology, Hanoi, Vietnam
(e-mail: \{binhmai2205, k100iltqbao, dinhhung15082004\}@gmail.com,
congtt@ptit.edu.vn)
}
\thanks{Corresponding author: Cong Tran (e-mail: congtt@ptit.edu.vn).}
}

\maketitle

\begin{abstract}
Diffusion-based text-to-audio (TTA) models achieve impressive synthesis quality but suffer from high inference latency due to iterative multi-step denoising. Existing one-step approaches alleviate this issue but still rely on paired text--audio data during distillation. To address these limitations, we propose SwiftAudio, a one-step TTA framework that performs audio-free distillation from a pretrained diffusion teacher using only text captions. Specifically, we adapt Variational Score Distillation (VSD) to the audio domain and introduce a temporal smoothness regularization objective to encourage coherent latent audio representations. This design enables the student model to inherit the teacher's generative prior without requiring paired audio supervision and allows effective training with only approximately 45K captions. Experiments on AudioCaps and Clotho demonstrate that SwiftAudio achieves state-of-the-art performance among strict one-step methods and substantially narrows the gap to multi-step diffusion systems. Project page: \url{https://swiftaudio.org/}
\end{abstract}


\section{Introduction}

\IEEEPARstart{T}{ext-to-Audio} (TTA) generation \cite{huang2023make, kreuk2023audiogen, liu23audioldm} has progressed rapidly in recent years, with diffusion-based models \cite{liu2024audioldm2, tang2023any, xue2024auffusion, majumder2024tango2} emerging as the dominant paradigm for synthesizing high-fidelity audio from natural language descriptions. Despite their impressive performance, diffusion models require iterative denoising over many sampling steps \cite{ho2020denoising, Song21denoising}, resulting in substantial inference latency and computational cost. This limitation hinders their deployment in real-time and resource-constrained applications.

\begin{figure}[t]
\centering
\includegraphics[width=\linewidth]{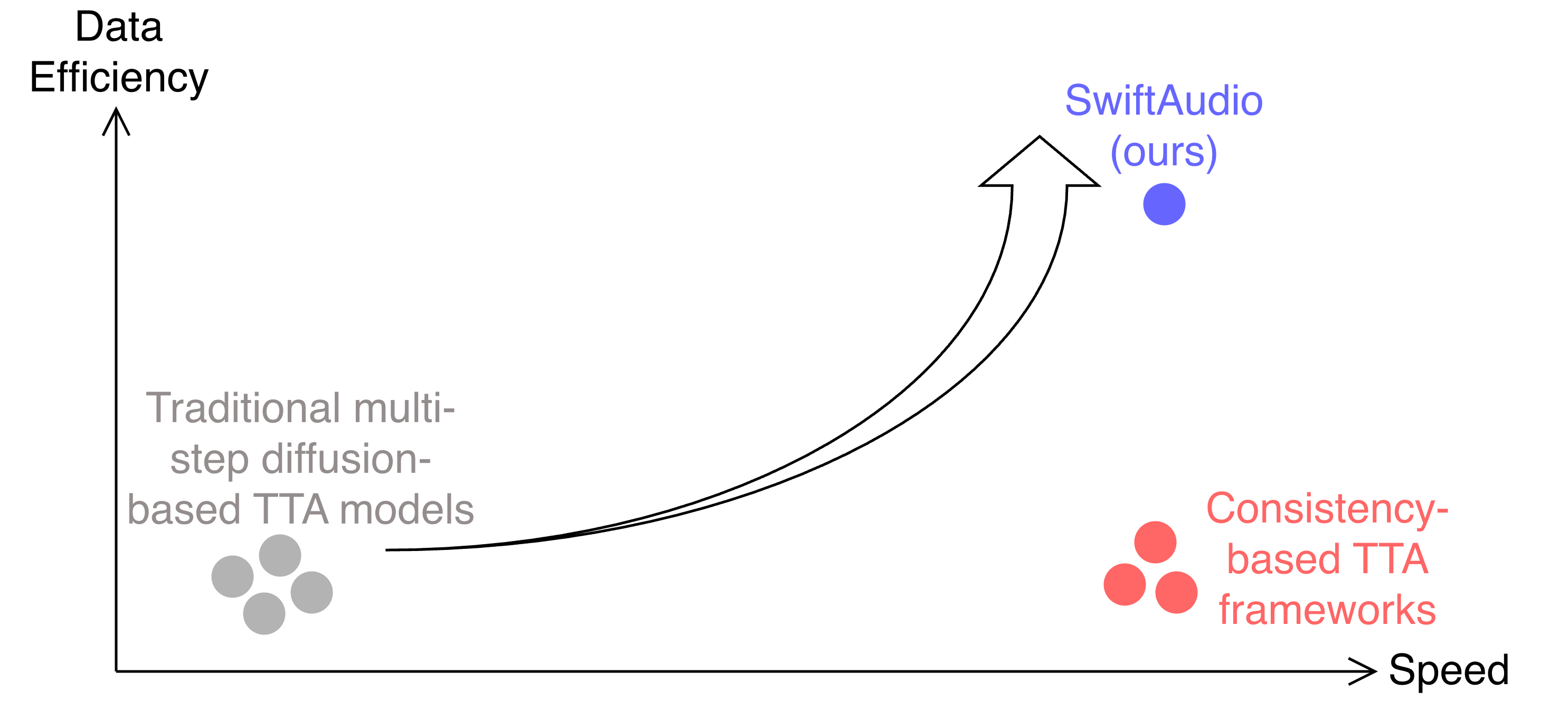}
\caption{
Conceptual illustration of existing TTA paradigms with respect to inference efficiency and data requirements. SwiftAudio advances along both dimensions by enabling one-step generation while eliminating the need for paired text--audio data during distillation.
}
\label{fig:compare}
\end{figure}

Recently advanced solvers \cite{Song21denoising,lu2022dpmsolver,lu2025dpm++,karras2022edm,baoanalytic} cut DDPM's \cite{ho2020denoising} sampling overhead to a few dozen steps; however, stable generation with fewer than 10 steps remains difficult to achieve. Consistency Models (CMs) \cite{song2023consistency} instead learn a direct mapping along the probability flow trajectory, enabling significantly faster sampling. Recent works extend this idea to TTA: ConsistencyTTA \cite{bai2024consistencytta} distills diffusion models into one-step generators, while AudioLCM \cite{liu2024audiolcm} reduces inference to two steps, and even one step, with minimal quality loss. Nevertheless, existing consistency-based TTA methods still exhibit two major limitations. First, generation quality degrades noticeably under strict one-step inference, especially for AudioLCM \cite{liu2024audiolcm}. Second, despite being distilled from pretrained teachers, both methods still require paired text–audio data during training.

{
Recent advances in Large Language Models (LLMs) and Vision Language Models (VLMs) have dramatically reduced the cost of obtaining textual descriptions. Captions can now be automatically expanded through LLM-based rewriting pipelines \cite{majumder2024tango2, brooks2023instructpix2pix} or generated directly from visual content by modern multimodal models \cite{MAI2025}. Consequently, large collections of diverse text prompts can be constructed with minimal human effort. In contrast, paired audio--caption datasets remain scarce and expensive to curate. AudioCaps \cite{kim2019audiocaps}, one of the few manually annotated benchmarks, contains only around 45K training examples, while larger resources such as WavCaps \cite{mei2023wavcaps} and AudioSetCaps \cite{bai2024audiosetcaps} rely heavily on metadata- or model-generated captions that may not faithfully describe acoustic events. As a result, high-quality audio--language corpora remain substantially smaller than modern image--language datasets containing millions or even billions of text annotations \cite{sun2023journeydb, schuhmann2022laion}. This discrepancy motivates an important question: \emph{Can a high-quality one-step text-to-audio generator be learned using only text captions and a pretrained diffusion teacher, without requiring paired caption--audio during distillation?}
}

However, achieving such a paradigm is non-trivial. Without paired audio supervision, the student must acquire both semantic alignment and audio generation capability solely from the teacher's learned distribution, while avoiding the quality degradation commonly observed in extreme one-step generation settings. To address these challenges, we introduce \textbf{SwiftAudio}, a one-step text-to-audio framework that enables \emph{audio-free distillation} from a pretrained diffusion teacher. Our approach adapts Variational Score Distillation (VSD) \cite{wang2023prolificdreamer} to transfer the teacher's generative prior using only text prompts, eliminating the need for paired audio-caption data during training. To improve generation quality under extreme one-step inference, we further propose a temporal-aware latent regularization objective that promotes coherent audio structures and complements the score-distillation signal. Together, these components enable inference-efficient and data-efficient text-to-audio generation, reducing inference to a single forward pass while removing the dependency on paired audio supervision. As illustrated in Figure~\ref{fig:compare}, SwiftAudio simultaneously improves inference efficiency and data efficiency, reducing generation to a single forward pass while removing the requirement for paired caption--audio during training.

{
Remarkably, we find that this caption-only distillation paradigm is highly data-efficient. While existing VSD-based image-free diffusion distillation methods in the visual domain \cite{nguyen2024swiftbrush, dao2024swiftbrush} typically require millions of prompts to train a one-step generator, SwiftAudio achieves strong performance using only approximately 45K captions from AudioCaps \cite{kim2019audiocaps}, where only the textual descriptions are retained and no paired audio examples are used during distillation.
}

Extensive experiments on AudioCaps \cite{kim2019audiocaps} and Clotho \cite{drossos2020clotho} demonstrate that SwiftAudio achieves state-of-the-art performance among strict one-step text-to-audio models while remaining competitive with substantially more expensive multi-step diffusion systems. Furthermore, our analyses reveal that the distilled generator preserves strong semantic controllability, shedding light on the representational properties of one-step audio generation models.

Our main contributions are summarized as follows:

\begin{itemize}

\item We propose \textbf{SwiftAudio}, a one-step text-to-audio generation framework that distills a pretrained diffusion teacher using only text captions, eliminating the need for paired audio--caption data during distillation.

\item We adapt Variational Score Distillation to the text-to-audio domain and combine it with a temporal-aware latent regularization objective, providing an audio-specific inductive bias for stable and coherent one-step generation.

\item We provide empirical evidence that caption-only distillation can be effective in the text-to-audio setting, with SwiftAudio achieving strong one-step generation performance using only approximately 45K captions, despite the limited scale of available audio-caption datasets.

\item Extensive experiments on AudioCaps and Clotho show that SwiftAudio achieves state-of-the-art performance among strict one-step text-to-audio models and substantially narrows the quality gap to multi-step diffusion systems.

\end{itemize}

\section{Related Work}

\subsection{Text-to-Audio Generation}

Text-to-audio (TTA) generation aims to synthesize audio signals conditioned on natural language descriptions and has advanced rapidly in recent years \cite{kreuk2023audiogen,liu23audioldm,liu2024audioldm2,xue2024auffusion}. Prior to the widespread adoption of continuous diffusion models, discrete representation modeling was a prevalent paradigm for neural audio generation. AudioLM \cite{Borsos2023AudioLM} demonstrated the effectiveness of hierarchical semantic tokens for high-quality audio continuation and unconditional synthesis. Building upon discrete token representations, AudioGen \cite{kreuk2023audiogen} formulated text-to-audio generation as autoregressive token prediction, while DiffSound \cite{yang2023diffsound} explored discrete diffusion probabilistic models for synthesizing audio from textual descriptions.

More recently, continuous diffusion models have emerged as the dominant paradigm for high-fidelity text-to-audio synthesis \cite{ho2020denoising,Song21denoising,karras2020score}. Representative systems include Make-An-Audio \cite{huang2023make}, AudioLDM \cite{liu23audioldm}, AudioLDM2 \cite{liu2024audioldm2}, Auffusion \cite{xue2024auffusion}, and Tango variations \cite{GhosalMMP23,majumder2024tango2}. To reduce the computational cost of waveform-level diffusion models such as DiffWave \cite{kong2021diffwave} and WaveGrad \cite{chen2021wavegrad}, most modern TTA systems adopt latent diffusion operating in compressed representation spaces \cite{rombach2022ldm}, as exemplified by AudioLDM \cite{liu23audioldm}. Despite their strong synthesis quality, these diffusion-based approaches rely on multi-step iterative denoising during inference, resulting in high latency and computational overhead.

\subsection{Fast Sampling for Diffusion Models}

The reliance on multi-step iterative denoising in diffusion models has motivated extensive research on fast sampling.
Solver-based approaches \cite{Song21denoising, lu2022dpmsolver,karras2022edm,lu2025dpm++,liupseudo,baoanalytic} accelerate sampling by reducing the number of inference steps, but still require multiple network evaluations.

Another line of work focuses on diffusion distillation, including Progressive Distillation~\cite{salimans2022progressive} and Consistency Models~\cite{song2023consistency}, which enable one-step generation by training a student model to approximate a multi-step teacher.
The latter method has been extended to TTA synthesis with ConsistencyTTA~\cite{bai2024consistencytta} and AudioLCM~\cite{liu2024audiolcm}, achieving significant speedups but still relying on paired text--audio data and often suffering quality degradation under strict one-step inference \cite{liu2024audiolcm}. Data-free diffusion distillation has recently shown promise in vision tasks \cite{nguyen2024swiftbrush, wang2023prolificdreamer}, and we discuss it in the next subsection.

\subsection{Score Distillation Methods}
Score Distillation Sampling (SDS) \cite{Poole23dreamfusion} marked a key advance in diffusion models by enabling the optimization of a target representation (e.g., 3D NeRF) through distillation from a pretrained 2D diffusion model, without ground-truth data. Although SDS enabled early text-to-3D methods \cite{Poole23dreamfusion, Wang2023sjc, Lin2023magic3d, Metzer2023latentnerf, chen2023fantasia3d}, it often suffers from over-saturation, over-smoothing, and limited diversity \cite{Poole23dreamfusion}. To mitigate these issues, Variational Score Distillation (VSD) \cite{wang2023prolificdreamer} models the target as a distribution rather than a single point, maintaining multiple particles and using a LoRA-adapted teacher model \cite{hu2022lora} to estimate variational scores. This formulation significantly improves fidelity and diversity over SDS. SwiftBrush \cite{nguyen2024swiftbrush} further extends VSD to image-free distillation for text-to-image generation, where a multi-step diffusion teacher is distilled into a one-step student generator using only text prompts, achieving competitive quality with substantially reduced inference cost.

Inspired by this line of work, SwiftAudio adapts VSD and SwiftBrush to the audio domain, enabling high-quality one-step TTA synthesis without heavy reliance on paired caption--audio training data. Table~\ref{tab:comparison_work} highlights the relationship between prior prevalent one-step distillation methods and our proposed approach.

\begin{table}[t]
\centering
\caption{
Comparison of representative one-step generation frameworks.
T2I denotes text-to-image generation and TTA denotes text-to-audio generation.
}
\label{tab:comparison_work}
\begin{tabular}{lccc}
\toprule
Method & Domain & Inference Steps & Distillation Data \\
\midrule
SwiftBrush \cite{nguyen2024swiftbrush} & T2I & 1 & Text-only \\
ConsistencyTTA \cite{bai2024consistencytta} & TTA & 1 & Text--Audio Pairs \\
AudioLCM \cite{liu2024audiolcm} & TTA & 1 & Text--Audio Pairs \\
\midrule
\textbf{SwiftAudio (Ours)} & \textbf{TTA} & \textbf{1} & \textbf{Text-only} \\
\bottomrule
\end{tabular}
\end{table}

\section{Methodology}

\begin{figure*}[t]
  \centering
  \includegraphics[width=\textwidth]{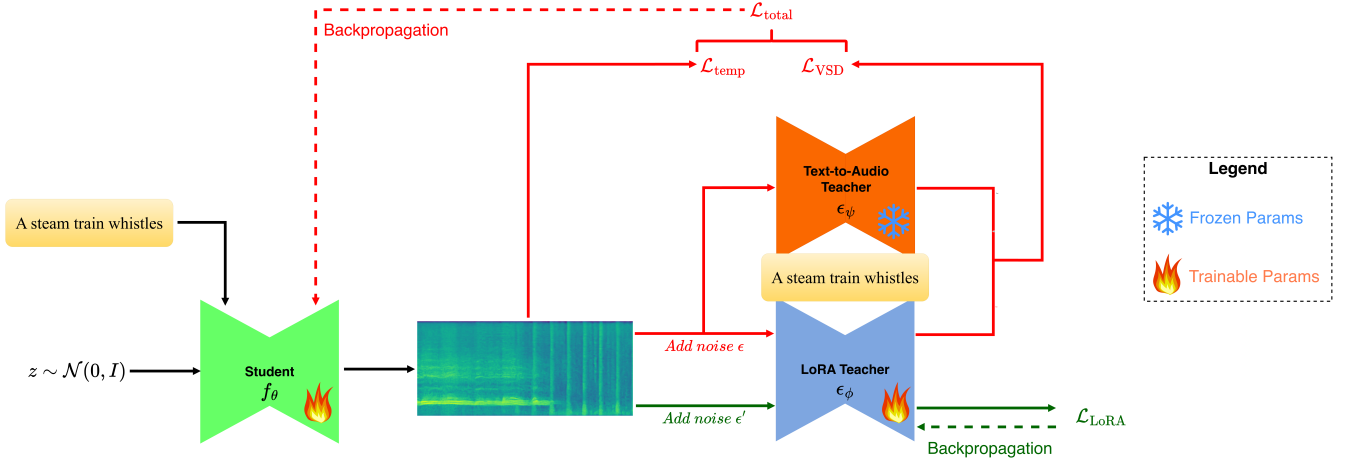}
  \caption{\textbf{Overview of the proposed SwiftAudio framework.} The student $f_\theta$ is trained using the total joint loss $\mathcal{L}_{\text{total}}$, integrating the VSD guidance ($\mathcal{L}_{\text{VSD}}$) from the teacher models and the temporal smoothness constraint ($\mathcal{L}_{\text{temp}}$) on the synthesized latents. The LoRA teacher $\epsilon_\phi$ is alternately updated via $\mathcal{L}_{\text{LoRA}}$ to accurately estimate the student score.}
  \label{fig:training}
\end{figure*}

\subsection{Overview}

SwiftAudio distills a pretrained multi-step text-to-audio diffusion model into a \emph{one-step} generator using only text captions, without requiring paired text--audio data during distillation. Our framework adapts Variational Score Distillation (VSD) \cite{wang2023prolificdreamer} to transfer the teacher's generative prior directly in latent space. To better capture the temporal nature of audio, we further introduce a temporal smoothness regularization term that encourages locally coherent latent representations. As illustrated in Figure~\ref{fig:training}, the student is optimized using a joint objective that combines VSD guidance and temporal regularization, while a LoRA-adapted teacher \cite{hu2022lora} is alternately updated to estimate the student score distribution. This design enables efficient one-step text-to-audio generation while eliminating the need for paired audio--caption supervision during distillation.

\subsection{Diffusion Setup and Notation}

Let $T$ denote the total number of diffusion timesteps and $\{\alpha_t, \sigma_t\}_{t=1}^{T}$ the predefined noise schedule. Both the student and the LoRA teacher use the same scheduler as the frozen teacher. During distillation, we assume access only to a text prompt dataset $\mathcal{D}=\{y\}$ without paired audio--caption samples. It is noted that while no ground-truth audio is used for student distillation, our proposed method inherits the generative prior of the pretrained teacher model, which was trained on paired audio--text data. 

\subsection{Model Components}

\textbf{Student ($f_\theta$).}
A one-step text-to-audio generator that maps Gaussian noise $z$ and text
conditioning $y$ to a clean audio latent $\hat{x}_0$.
We implement $f_\theta$ via a diffusion-style noise predictor network and
a deterministic reparameterization (details in Sec.~\ref{sec:parameterization}).

\textbf{Frozen Teacher ($\epsilon_\psi$).}
A pretrained multi-step diffusion model that remains fixed during training and
provides a high-quality generative prior.

\textbf{LoRA Teacher ($\epsilon_\phi$).}
A lightweight LoRA-adapted version of the teacher, trained to estimate the score
of the student distribution and provide corrective gradients for VSD.

\subsection{One-step Student Parameterization}\label{sec:parameterization}

We aim to learn a one-step generator that maps Gaussian noise to a clean
audio latent. Let $z \sim \mathcal{N}(0,I)$ denote the noisy latent at the
final diffusion timestep $T$ under the schedule defined above.

We parameterize the generator $f_\theta$ via a diffusion-style noise predictor
network, following \cite{nguyen2024swiftbrush}:
\begin{equation}
\hat{x}_0 = f_\theta(z,y)
=
\frac{z - \sigma_T \, \epsilon_\theta(z, T, y)}{\alpha_T},
\label{eq:student_param}
\end{equation}
where $\epsilon_\theta$ predicts the noise component at time $T$.
In implementation, the learnable network is $\epsilon_\theta$, while $f_\theta$
denotes the induced one-step mapping used for sampling.

\subsection{Audio-Free Distillation Mechanism}

\subsubsection{Student Update}
\paragraph{VSD Guidance}
With the student generator parameterized as above, we now describe
how it is optimized without ground-truth audio using score-based distillation.
We view the student $f_\theta$ as defining an implicit distribution over audio
latents and aim to minimize its Kullback–Leibler (KL) divergence to the teacher prior.
Since the KL objective is intractable, we adopt VSD \cite{wang2023prolificdreamer} to estimate its gradient. 

Specifically, given a student-generated latent $\hat{x}_0 = f_\theta(z,y)$, we
sample an intermediate noisy state 
\begin{equation}
x_t = \alpha_t \hat{x}_0 + \sigma_t \epsilon,
\end{equation}
where $\epsilon \sim \mathcal{N}(0, I)$ and $t \sim \mathcal{U}(0.02T, 0.98T)$, following prior work \cite{Poole23dreamfusion, nguyen2024swiftbrush, wang2023prolificdreamer}.

The student is then updated using the score difference between the frozen teacher
and the LoRA teacher:

\begin{equation}
\begin{split}
\nabla_{\theta}\mathcal{L}_{\text{VSD}} = \mathbb{E}_{t,\epsilon,y} \bigg[ & \omega(t)(\epsilon_{\psi}(x_{t},t,y)-\epsilon_{\phi}(x_{t},t,y)) \\
& \times \frac{\partial f_\theta(z,y)}{\partial\theta} \bigg],
\end{split}
\label{eq:vsd}
\end{equation}
where $\omega(t)$ is a weighting function.

{
\paragraph{Temporal Regularization}

\begin{figure*}[t]
\centering
\includegraphics[width=\textwidth]{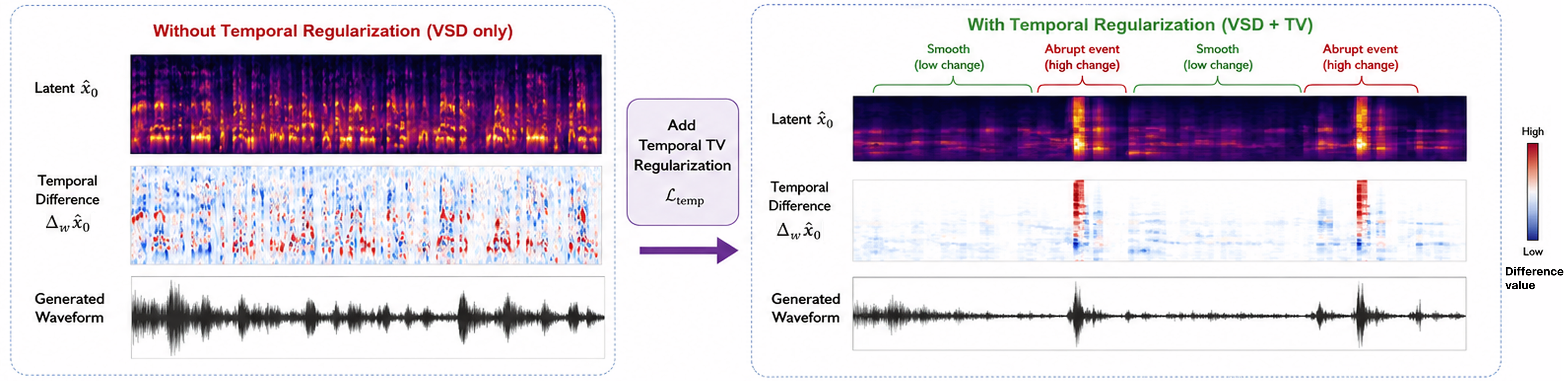}
\caption{
Conceptual illustration of temporal regularization. 
Compared with VSD-only generation, adding temporal TV encourages a piecewise-smooth latent trajectory by suppressing spurious frame-to-frame fluctuations, while still allowing localized temporal changes when required by abrupt acoustic events.
}
\label{fig:temporal_regularization}
\end{figure*}

While the VSD objective aligns the student distribution with that of the pretrained teacher, it does not explicitly constrain temporal consistency in the generated latent representation. As a result, one-step generation may exhibit unstable latent trajectories, particularly in acoustically complex scenes.

To encourage temporal coherence, we introduce a temporal Total Variation (TV) regularization term defined along the latent time dimension. Formally, for a continuous, differentiable function or signal, the Total Variation norm is defined as the integral of the magnitude of its gradient \cite{rudin1992nonlinear}:
\begin{equation}
\text{TV}(f) = \int |\nabla f(t)| dt.
\end{equation}
When mapping this formulation to a discrete and multi-dimensional grid, such as the generated latent representation $\hat{x}_0 \in \mathbb{R}^{C \times F \times W}$ (where $C$, $F$, and $W$ denote the channel, frequency, and temporal dimensions, respectively), the continuous gradient along the time axis is approximated via finite differences. Specifically, the first-order backward temporal difference at time step $w$ is expressed as:
\begin{equation}
\Delta_{w} \hat{x}_{0}(\cdot,\cdot,w) = \hat{x}_{0}(\cdot,\cdot,w) - \hat{x}_{0}(\cdot,\cdot,w-1),
\end{equation}
where $\hat{x}_{0}(\cdot,\cdot,w)$ denotes the latent slice at the w-th temporal position, retaining all channel and frequency dimensions.

By substituting this finite difference approximation into the continuous TV definition and discretizing the integral over the temporal grid, the penalty naturally takes the form of the $L_1$ norm of the temporal transitions \cite{zach2007duality}. Its effectiveness stems precisely from this $L_1$ norm formulation, which is well-known for promoting sparsity in the gradient domain.

In the temporal domain, most acoustic content evolves smoothly over time, whereas abrupt events such as dog barks, thunder strikes, door slams, percussive onsets, and phonetic boundaries occur sparsely. Intuitively, the sparsity-inducing nature of the TV regularizer encourages smooth latent evolution in stationary regions while preserving a small number of significant temporal discontinuities. This behavior is analogous to edge preservation in classical image denoising and is particularly desirable for audio generation, where transient structures often carry important perceptual cues. Consequently, we define our temporal regularization loss $\mathcal{L}_{\text{temp}}$ as:
\begin{equation}
\begin{split}
\mathcal{L}_{\mathrm{temp}} = \mathbb{E}_{z,y} \bigg[ & \frac{1}{C F (W-1)} \\
& \times \sum_{c=1}^{C} \sum_{f=1}^{F} \sum_{w=2}^{W} \left| \hat{x}_{0}(c,f,w) - \hat{x}_{0}(c,f,w-1) \right| \bigg].
\end{split}
\end{equation}
This regularizer encourages piecewise-smooth latent evolution over time, thereby reducing spurious frame-to-frame instability. 
Because the penalty is based on an $L_1$ norm of temporal differences, it tends to concentrate temporal changes into sparse locations rather than enforcing uniform smoothness everywhere. 
When combined with the VSD objective, this allows meaningful abrupt acoustic events to remain localized while stationary regions evolve more smoothly. 
As conceptually illustrated in Fig.~\ref{fig:temporal_regularization}, temporal regularization suppresses unstable temporal fluctuations in smooth regions while permitting localized changes around salient acoustic events.

\paragraph{Total Objective Function}
To simultaneously leverage the generative prior of the teacher model and ensure the structural smoothness of the generated audio latents, we combine the VSD distillation gradient with the temporal regularization loss. The total joint objective function for optimizing the student network is defined as:
\begin{equation}
\mathcal{L}_{\text{total}} = \lambda \cdot \mathcal{L}_{\text{temp}} + \mathcal{L}_{\text{VSD}},
\label{eq:total_loss}
\end{equation}
where $\lambda$ represents a hyperparameter that balances the weight of the temporal smoothness constraint against the score distillation objective. This optimization corresponds to \textbf{Phase 2} in Algorithm~\ref{alg:distillation}.

}

\subsubsection{LoRA Teacher Update}

Since the score function of the implicit student distribution is unavailable in closed form, we approximate it using a LoRA-adapted teacher. Following \cite{wang2023prolificdreamer, nguyen2024swiftbrush}, we approximate it by training a LoRA-adapted teacher
$\epsilon_\phi$ on student-generated samples.

Concretely, $\epsilon_\phi$ is optimized to predict the noise added to
$\hat{x}_0$, using the standard diffusion denoising objective:

\begin{equation}
\mathcal{L}_{\text{LoRA}}
=
\mathbb{E}_{t',\epsilon',y}\Bigl[
\bigl\lVert
\epsilon_{\phi}\left(
\alpha_{t'} \hat{x}_0 + \sigma_{t'} \epsilon',\, t',\, y
\right)
- \epsilon'
\bigr\rVert_2^2
\Bigr],
\label{eq:lora}
\end{equation}
where $\epsilon' \sim \mathcal{N}(0,I)$ and $t' \sim \mathcal{U}(0, T)$. 

This step corresponds to \textbf{Phase 3} in Algorithm~\ref{alg:distillation},
allowing the LoRA teacher to track the evolving student distribution and provide
accurate VSD gradients.

\subsection{Training and Inference}

Training alternates between student updates and LoRA teacher adaptation,
as summarized in Algorithm~\ref{alg:distillation}. At inference time, the
multi-step diffusion process is collapsed into a single forward pass
(Algorithm~\ref{alg:infer}), enabling extremely fast audio generation.

\begin{algorithm}[!t]
\caption{Audio-free Distillation Training (Pytorch-like Pseudocode)}
\label{alg:distillation}
\begin{algorithmic}[1]
\Require $\epsilon_\psi$ (frozen); $\epsilon_\phi$; $f_\theta$; learning rates $\eta_1$ and $\eta_2$; balancing weight $\lambda$; $\mathcal{D}$; $T$; $\{\alpha_t, \sigma_t\}_{t=1}^T$; $\omega$.
\State Initialize $\phi \leftarrow \psi, \theta \leftarrow \psi$

\While{not converged}
    \State Sample inputs: $y \sim \mathcal{D}$, $z \sim \mathcal{N}(0, I)$
    
    \Statex \hskip\algorithmicindent {\textbf{\# Phase 1: Student Generation}}
    \State Compute $\hat{x}_0 = f_\theta(z, y)$
    
    \Statex \hskip\algorithmicindent {\textbf{\# Phase 2: Joint Guidance \& Regularization (Student Update)}}
    \State Sample $t \sim \mathcal{U}(0.02T, 0.98T)$, $\epsilon \sim \mathcal{N}(0, I)$
    \State $x_t \leftarrow \alpha_t \hat{x}_0 + \sigma_t \epsilon$ 
    
    \With{no\_grad}
        \State $\delta_{\text{score}} \leftarrow \omega(t) \cdot (\epsilon_\psi(x_t, t, y) - \epsilon_\phi(x_t, t, y))$
        \State $x_{\text{target}} \leftarrow \hat{x}_0 - \delta_{\text{score}}$ 
    \EndWith
    
    \State $\mathcal{L}_{\text{VSD}} \leftarrow \frac{1}{2} || \hat{x}_0 - x_{\text{target}} ||^2$ 
    \State $\mathcal{L}_{\text{temp}} \leftarrow
    \operatorname{mean}
    \left(
    \left|
    \hat{x}_{0}[:,:,1:]
    -
    \hat{x}_{0}[:,:,:-1]
    \right|
    \right)$
    \State $\mathcal{L}_{\text{total}} \leftarrow \lambda \cdot \mathcal{L}_{\text{temp}} + \mathcal{L}_{\text{VSD}}$
    \State Update $\theta \leftarrow \theta - \eta_1 \nabla_\theta \mathcal{L}_{\text{total}}$
    
    \Statex \hskip\algorithmicindent {\textbf{\# Phase 3: LoRA Adaptation (Teacher Update)}}
    \State Sample $t' \sim \mathcal{U}(0, T)$, $\epsilon' \sim \mathcal{N}(0, I)$
    \State $x'_{t'} \leftarrow \alpha_{t'} \cdot \text{stop\_grad}(\hat{x}_0) + \sigma_{t'} \epsilon'$ 
    
    \State $\mathcal{L}_{\text{LoRA}} \leftarrow || \epsilon_\phi(x'_{t'}, t', y) - \epsilon' ||^2$
    \State Update $\phi \leftarrow \phi - \eta_2 \nabla_\phi \mathcal{L}_{\text{LoRA}}$
\EndWhile
\end{algorithmic}
\end{algorithm}

\begin{algorithm}[!t]
\caption{One-step Sampling}
\label{alg:infer}
\begin{algorithmic}[1]
\Require SwiftAudio $f_\theta$; final timestep T; text prompt $y$; VAE decoder $D(\cdot)$; vocoder $V(\cdot)$.
\State Sample $z \sim \mathcal{N}(0, I)$
\State Compute $\hat{x}_0 = f_\theta(z, y)$ 
\State \Return $V(D(\hat{x}_0))$
\end{algorithmic}
\end{algorithm}

\section{Experiments}

\subsection{Experimental Setup}

\textbf{Implementation details.}
We distill the student model $f_\theta$ using two teacher networks, $\epsilon_\psi$ and $\epsilon_\phi$. During training, both teachers generate denoising predictions under classifier-free guidance (CFG) \cite{ho2021classifier} with a guidance scale of $7.5$. The student model and both teacher models are initialized from the same pretrained Auffusion checkpoint \cite{xue2024auffusion}\footnote{\url{https://huggingface.co/auffusion/auffusion-full-no-adapter}}. Following Auffusion, we use latent audio representations of size $C \times F \times W = 4 \times 32 \times 128$, corresponding to the channel, frequency, and temporal dimensions, respectively.

The student model is optimized using AdamW \cite{loshchilov19adamw} with a learning rate of $1\times10^{-5}$. For the trainable teacher branch, we adopt a parameter-efficient fine-tuning strategy based on LoRA \cite{hu2022lora}, where only the LoRA parameters are updated. The LoRA teacher is optimized with AdamW using a learning rate of $1\times10^{-3}$, a rank of $r=64$, and a scaling factor $\alpha=128$.

We set the temporal regularization coefficient to $\lambda = 0.05$. The final-step coefficients used in Eq.~(\ref{eq:student_param}) are $\alpha_T = 0.9953^{0.5}$ and $\sigma_T = 0.0047^{0.5}$.

Following the formulation in Eq.~(\ref{eq:vsd}), we set the weighting function to $\omega(t)=\sigma_t^2$. Training is conducted for 20,000 optimization steps with an effective batch size of 64, achieved using a per-device batch size of 16 and gradient accumulation over 4 steps. All experiments are performed on a single NVIDIA RTX 5880 Ada GPU with 48GB of VRAM. Under this setup, the complete training process requires approximately 40 hours.

\textbf{Dataset.} We use the text captions from the AudioCaps dataset \cite{kim2019audiocaps} as the sole source of training data. AudioCaps is currently the largest publicly available human-annotated audio captioning dataset, providing high-quality textual descriptions for a wide range of audio events. In our experiments, we train on approximately 45K captions from the training split and evaluate on around 750 samples from a subset of the test split, excluding samples that are unavailable due to copyright restrictions.

Despite its scale within the audio captioning domain, AudioCaps remains relatively small compared to image-language datasets \cite{sun2023journeydb, schuhmann2022laion}, which often contain millions of captioned examples. The limited size of AudioCaps therefore presents an additional challenge for training text-to-audio generation models solely from paired caption data.

Following \cite{xue2024auffusion}, we further evaluate the out-of-domain generalization ability of our model through zero-shot experiments on the Clotho dataset \cite{drossos2020clotho}. 

\textbf{Evaluation metrics.} We employ FAD, FD, IS, and KL for objective evaluation, following prior works \cite{liu23audioldm, xue2024auffusion}. FAD, KL and IS are computed using PANN embeddings \cite{kong2020panns}, while FD is based on VGGish embeddings \cite{hershey2017vggish}. The evaluation codebase is provided in the following project\footnote{\url{https://github.com/haoheliu/audioldm_eval}}.

For subjective evaluation, we conduct a Mean Opinion Score (MOS) study with 31 participants, who rate overall quality (OVL) and text relevance (REL) on a 5-point scale; participants provided informed consent, no personally identifiable information was collected, and formal ethics review was not required for anonymous perceptual ratings. Detailed protocols and interfaces are provided in the Supplementary Material. 

\textbf{Baselines and evaluation protocol.}
We compare SwiftAudio with currently state-of-the-art diffusion-based audio generation models, including multi-step (AudioLDM2 \cite{liu2024audioldm2}, Auffusion~\cite{xue2024auffusion}) and fast distilled methods (AudioLCM \cite{liu2024audiolcm}, ConsistencyTTA \cite{bai2024consistencytta}). Multi-step baselines use standard sampling with CFG, while fast methods are evaluated in the strict one-step setting.

Following \cite{bai2024consistencytta}, \#Queries counts the total number of denoising network evaluations. All multi-step methods use CFG, so each step requires one conditional and one unconditional pass (i.e. 200 queries = 100 unconditional steps + 100 conditional steps). AudioLCM and ConsistencyTTA use a single conditional pass without CFG at inference time. 

For reproducibility, checkpoints for evaluations are listed in the Supplementary Material.

\subsection{Experimental Results}

\subsubsection{Performance on AudioCaps}

\begin{table*}[t]
    \centering
    \caption{Comparison of different methods on the AudioCaps dataset. Methods are grouped into \textbf{multi-step} vs. \textbf{one-step} for fair comparison. \textbf{Duration} denotes the total audio duration of the \emph{training set} used by each method (hours). \textbf{\#Queries} is the number of denoising network evaluations at inference. Best results are highlighted \textbf{within each block}.}
    \label{tab:comparison}

    \resizebox{1\textwidth}{!}{%
    \begin{tabular}{l lcc cccc cc}
        \toprule
        \multirow{2}{*}{\textbf{Type}} &
        \multirow{2}{*}{\textbf{Method}} &
        \multirow{2}{*}{\textbf{Duration (h)}} &
        \multirow{2}{*}{\textbf{\#Queries}} &
        \multicolumn{4}{c}{\textbf{Objective metric}} &
        \multicolumn{2}{c}{\textbf{Subjective metric}} \\
        
        \cmidrule(lr){5-8}\cmidrule(lr){9-10}
        & & & &
        \textbf{FD} $\downarrow$ &
        \textbf{FAD} $\downarrow$ &
        \textbf{KL} $\downarrow$ &
        \textbf{IS} $\uparrow$ &
        \textbf{OVL} $\uparrow$ &
        \textbf{REL} $\uparrow$ \\
        \midrule

        \multirow{2}{*}{Multi-step}
        & AudioLDM2 \cite{liu2024audioldm2} & 29,510 & 200 & 23.42 & \textbf{1.87} & 1.68 & 9.52 & 3.77 & 3.68 \\
        & Auffusion-full (Teacher) \cite{xue2024auffusion} & 1,990 & 200 & \textbf{22.49} & 1.91 & \textbf{1.43} & \textbf{10.42} & \textbf{4.06} & \textbf{4.10} \\
        
        \midrule

        \multirow{3}{*}{One-step}
        & AudioLCM \cite{liu2024audiolcm} & 110 & 1 & 23.15 & 2.92 & 1.75 & 5.81 & 3.26 & 3.55 \\
        & ConsistencyTTA \cite{bai2024consistencytta} & 110 & 1 & 25.68 & 3.37 & \textbf{1.42} & \textbf{9.26} & 3.74 & \textbf{3.94} \\
        & SwiftAudio (Ours) & None & 1 & \textbf{22.73} & \textbf{2.25} & 1.62 & 9.13 & \textbf{3.90} & 3.87 \\
        
        \bottomrule
    \end{tabular}}
\end{table*}

Table~\ref{tab:comparison} compares SwiftAudio with state-of-the-art multi-step and one-step TTA models on AudioCaps. Among strict one-step approaches, SwiftAudio achieves the best performance, obtaining lower FD (22.73) and FAD (2.25) than AudioLCM and ConsistencyTTA. Notably, it substantially narrows the gap to its multi-step teacher, Auffusion-full (FD: 22.73 vs.\ 22.49), while reducing inference to a single denoising query (200$\times$ fewer evaluations). Importantly, this is achieved without requiring additional paired text--audio data during distillation.

Remarkably, SwiftAudio even outperforms AudioLDM2 on FD while using a teacher trained on only 1,990 hours of audio, compared to the 29,510 hours required by AudioLDM2. Considering the substantial disparity in training data scale and associated computational cost, these results suggest that knowledge distillation from a high-quality latent diffusion teacher can be more effective than scaling training data alone for achieving high-fidelity one-step audio generation.

Subjective evaluations further show that SwiftAudio attains the highest overall quality (OVL) and competitive relevance (REL) scores among one-step models, although the teacher model remains superior. We further investigate the robustness of the learned generative prior through a zero-shot evaluation on the Clotho dataset, presented in the following subsection.

{
\subsubsection{Zero-shot Generalization on Clotho}

\begin{table*}[t]
    \centering
    \caption{Comparison of different one-step methods on the Clotho dataset. \textbf{Duration} denotes the total audio duration of the \emph{training set} used by each method (hours); SwiftAudio uses \emph{caption-only} training. \textbf{\#Queries} is the number of denoising network evaluations at inference.}

    \label{tab:zeroshot}
    \resizebox{.7\textwidth}{!}{%
    \begin{tabular}{lcc cccc}
        \toprule
        \textbf{Method} &
        \textbf{Duration (h)} &
        \textbf{\#Queries} &
        \textbf{FD} $\downarrow$ &
        \textbf{FAD} $\downarrow$ &
        \textbf{KL} $\downarrow$ &
        \textbf{IS} $\uparrow$\\
        \midrule
        AudioLCM & 110 & 1 & \textbf{23.18} & 4.42 & 2.54 & 6.38   \\
        ConsistencyTTA & 110 & 1 & 30.01 & 5.13 & 2.48 & 7.02  \\
        SwiftAudio (Ours) & None & 1 & 23.45 & \textbf{2.56} & \textbf{2.13} & \textbf{7.38} \\
        
        \bottomrule
    \end{tabular}
    }
\end{table*}

To evaluate out-of-domain generalization, we conduct zero-shot experiments on the Clotho dataset \cite{drossos2020clotho}. Since Clotho recordings are typically longer than the 10-second generation duration used by all compared methods, we randomly crop a 10-second segment from each evaluation audio. Each Clotho sample is associated with five human-written captions, not all of which necessarily describe the selected segment. To reduce this mismatch, we compute the CLAP similarity \cite{laionclap2023} between the cropped audio and its five reference captions, and use the most compatible caption as the generation prompt. Additional details are provided in the Supplementary Material.

Table~\ref{tab:zeroshot} reports the results. The results further reveal a substantial difference in cross-dataset robustness. Although ConsistencyTTA \cite{bai2024consistencytta} performs competitively on AudioCaps \cite{kim2019audiocaps}, its performance deteriorates considerably on Clotho, suggesting a stronger dependence on the training-domain distribution. In contrast, SwiftAudio exhibits substantially better transferability, indicating that the proposed caption-only distillation framework learns a more general text-conditioned generative prior rather than merely reproducing teacher behavior on AudioCaps.
}

\begin{figure}[t]
  \centering
  \includegraphics[width=\linewidth]{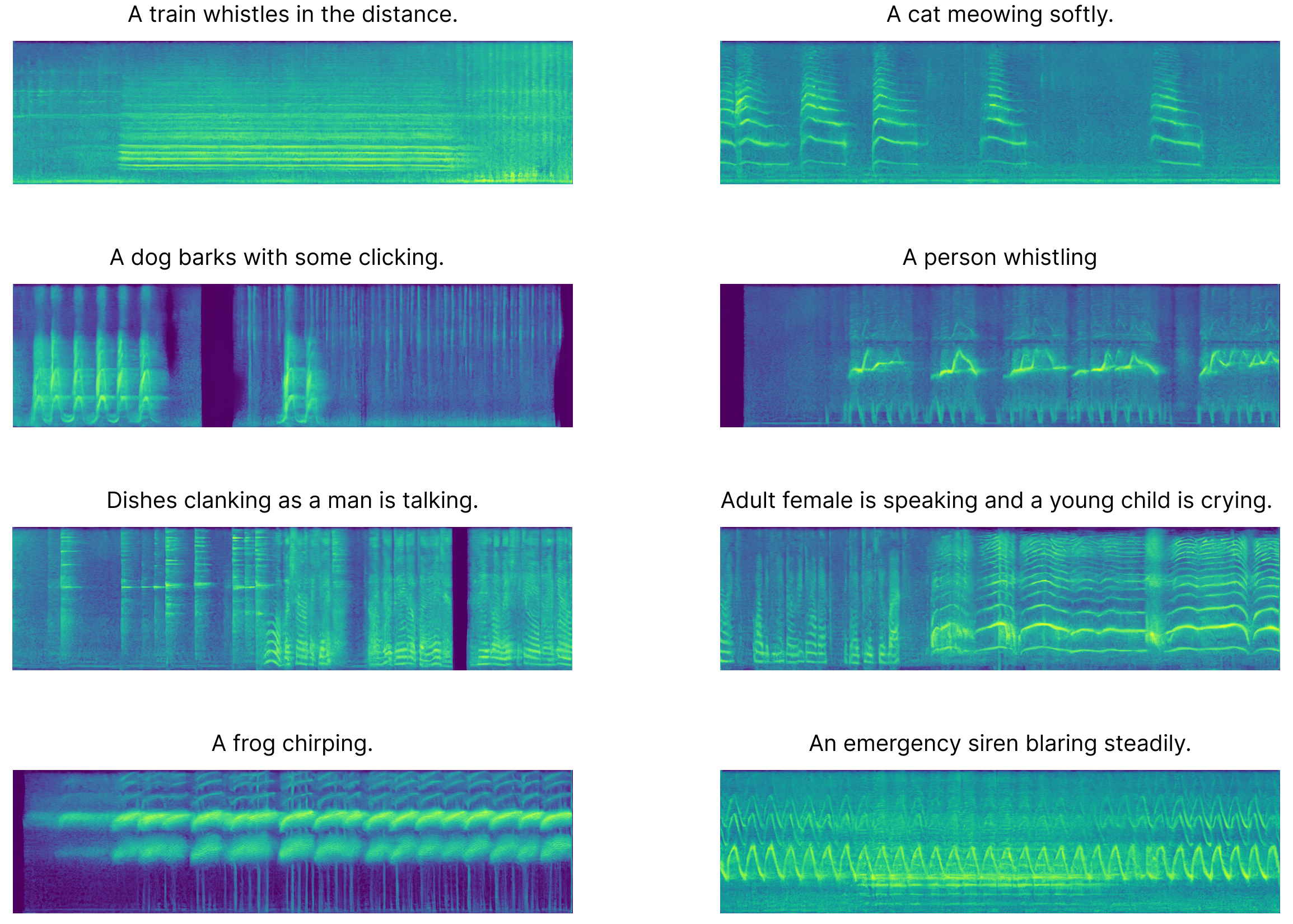}
  \caption{Qualitative results of the SwiftAudio framework. The figure illustrates the mel-spectrograms of generated audio samples corresponding to various text prompts, demonstrating the model's capability to synthesize detailed and diverse acoustic patterns in a single inference step.}
  \label{fig:qualitative}
\end{figure}

{
\subsubsection{Qualitative Evaluation}
To visually demonstrate the synthesis capabilities of our proposed method, Figure~\ref{fig:qualitative} presents mel-spectrograms of audio samples generated by SwiftAudio conditioned on diverse text prompts. As shown, the model captures distinct acoustic characteristics, ranging from transient events such as clanking dishes to continuous, structured harmonics such as an emergency siren. This result confirms that SwiftAudio maintains high text relevance and synthesis fidelity while requiring only a single denoising step.
}

\subsection{Analysis}

\subsubsection{Ablation Study}\label{sec:ablation}

\begin{table}[t]
\centering
\caption{Ablation results validating key design choices.}
\label{tab:ablation}
\resizebox{\linewidth}{!}{
\begin{tabular}{lcccc}
\toprule
\textbf{SwiftAudio} & \textbf{FD $\downarrow$} & \textbf{FAD $\downarrow$} & \textbf{KL $\downarrow$} & \textbf{IS $\uparrow$} \\
\midrule
w/o Student Parameterization
    & 47.13 & 8.73 & 3.36 & 4.58 \\

\midrule

w/ LoRA Teacher ($r=4$, $\alpha=8$)
    & 56.14 & 9.71 & 3.22 & 3.63 \\

w/ LoRA Teacher ($r=32$, $\alpha=64$)
    & 27.85 & 4.37 & 1.72 & 5.73 \\

\midrule
w/o Temporal Regularization
    & 23.19 & 3.47 & 1.67 & 8.04 \\

w/ L2 Temporal Regularization
    & 23.61 & 2.81 & \textbf{1.52} & 8.83 \\

\midrule
SwiftAudio (Proposed Method)
    & \textbf{22.73} & \textbf{2.25} & 1.62 & \textbf{9.13} \\

\bottomrule
\end{tabular}
}
\end{table}

We conduct ablation experiments to verify the effectiveness of our design choices, with results summarized in Table~\ref{tab:ablation}.

\textbf{Effect of one-step student parameterization.} Replacing the proposed diffusion-style parameterization (Eq.~\ref{eq:student_param}) with a naive direct mapping leads to a severe performance drop, with FD increasing from 22.73 to 47.13. This confirms that the noise-prediction formulation is essential for stabilizing the distillation process.

\textbf{Impact of LoRA teacher capacity.}

To isolate the role of the LoRA teacher in distillation, we fix the student architecture  and vary only the LoRA rank of the teacher. Reducing the rank to $r=4$ significantly degrades performance (FD 56.14), indicating that an under-parameterized teacher provides inaccurate supervision signals. Increasing the capacity ($r=32$) substantially improves all metrics. These results suggest that the quality of distillation is strongly bounded by the expressive power of the teacher used to generate score targets.

{
\textbf{Effect of temporal regularization.}

Table~\ref{tab:ablation} investigates the impact of temporal regularization on the distilled one-step generator. Removing the temporal regularization term results in inferior performance across most perceptual metrics, with FD increasing from 22.73 to 23.19 and FAD rising substantially from 2.25 to 3.47. The proposed model also achieves the highest Inception Score (9.13), indicating improved audio quality and diversity. Although the FD improvement is small, the consistent gains in FAD and IS suggest that temporal regularization provides an additional optimization signal that complements the Variational Score Distillation objective \cite{wang2023prolificdreamer} and encourages more coherent temporal generation.

We further compare the proposed temporal regularization with a conventional L2-based smoothness constraint. While the L2 variant achieves the lowest KL divergence, it produces worse FD and FAD scores than the proposed approach. This observation can be attributed to the different inductive biases imposed by the two penalties. Specifically, the L2 objective encourages uniformly smooth latent trajectories by heavily penalizing large temporal variations, which may inadvertently suppress transient structures and abrupt changes that are intrinsic to many audio events. In contrast, the proposed TV-inspired L1 regularization promotes piecewise-smooth temporal dynamics, reducing unnecessary fluctuations while preserving salient temporal transitions. Consequently, it achieves a more favorable trade-off between distributional alignment and perceptual quality, leading to the best overall performance among all temporal regularization variants.
}

{
\subsubsection{Discussion on Caption-Only Data Efficiency}

SwiftAudio demonstrates strong data efficiency in the caption-only distillation setting. Existing vision-based one-step prompt-only distillation studies are commonly conducted at much larger prompt scales; for example, SwiftBrush~\cite{nguyen2024swiftbrush} uses 1.38M text prompts. By comparison, SwiftAudio achieves competitive one-step text-to-audio performance using only approximately 45K human-written AudioCaps captions, about 30$\times$ fewer prompts. This comparison is intended to contextualize the scale of our training data, rather than to imply that image-domain distillation would fail with fewer prompts.

Table~\ref{tab:data_scaling} examines how distillation performance changes as the number of training captions increases. Overall, generation quality improves as more captions are used. The 5K and 10K settings lead to clearly weaker performance, while increasing the caption set to 20K yields a substantial improvement in FAD and IS. Using the full AudioCaps caption set further improves all metrics, suggesting that broader prompt coverage benefits both perceptual quality and distributional alignment.

Importantly, these results should not be interpreted as showing that 45K captions are sufficient in general, or that larger caption collections would not further improve performance. Rather, they indicate that SwiftAudio can already obtain strong one-step generation performance under a constrained but high-quality caption-only training regime. This setting is particularly relevant for text-to-audio generation, where manually verified audio captions remain much harder to obtain than text prompts in image generation.


One possible explanation for the observed data efficiency is that environmental audio captions often contain recurring sound-event concepts expressed through different textual variations. Such redundancy may allow the student to sample a useful portion of the teacher's conditional generation space even from a relatively small caption set. However, we view this as a hypothesis rather than a definitive conclusion, and leave a systematic investigation of modality-dependent and dataset-dependent scaling behavior to future work.

\begin{table}[t]
\centering
\caption{Impact of text prompt quantity on distillation performance. The models are distilled using different subsets of captions from the AudioCaps training split.}
\label{tab:data_scaling}
\resizebox{\linewidth}{!}{
\begin{tabular}{lcccc}
\toprule
\textbf{Caption Quantity} & \textbf{FD $\downarrow$} & \textbf{FAD $\downarrow$} & \textbf{KL $\downarrow$} & \textbf{IS $\uparrow$} \\
\midrule
5K samples & 36.41 & 6.45 & 1.88 & 4.61 \\
10K samples & 36.84 & 10.49 & 1.84 & 4.79 \\
20K samples & 32.30 & 2.98 & 1.84 & 7.56 \\
$\sim$45K (Full) & \textbf{22.73} & \textbf{2.25} & \textbf{1.62} & \textbf{9.13} \\
\bottomrule
\end{tabular}
}
\end{table}
}

{
\subsubsection{Analysis of Knowledge Transfer and Semantic Preservation}
\label{sec:semantic_control}

\begin{figure*}[t]
  \centering
  \includegraphics[width=\linewidth]{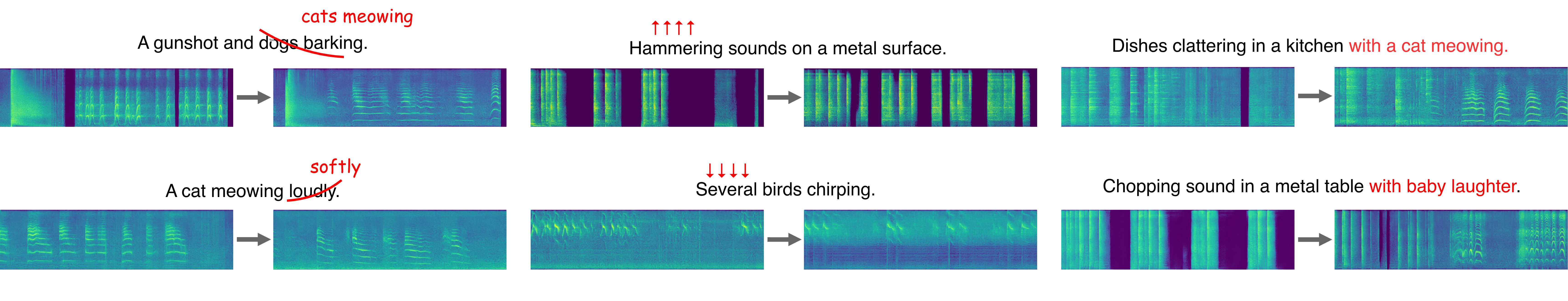} 
  \caption{Semantic Controllability and Latent Disentanglement in SwiftAudio. \textbf{(Right)} Word swapping modifies specific acoustic properties while maintaining scene consistency. \textbf{(Middle)} Attention reweighting modulates sound intensity and temporal density. \textbf{(Left)} Word refinement compositionally integrates new semantic elements without altering the base sound event.}
  \label{fig:semantic_control}
\end{figure*}

While recent one-step frameworks like NASA \cite{Nguyen_2025_ICCV} demonstrate binary feature-steering via negative prompts, we investigate whether SwiftAudio preserves a continuous and dynamic semantic space. Inspired by Prompt-to-Prompt \cite{hertz2023prompt2prompt}, rather than editing real audio signals, we manipulate the text conditioning under a fixed initial noise latent $z$ to verify if semantic concepts remain disentangled in the distilled deterministic mapping:

\begin{itemize}
    \item \textbf{Word swapping:} Replacing specific keywords (e.g., ``dogs barking'' to ``cats meowing'') accurately alters the target sound event while seamlessly preserving the unedited acoustic background.
    
    \item \textbf{Attention reweighting:} Modulating the attention weights of text tokens controls the spectral intensity and temporal density of specific sounds, demonstrating continuous semantic scaling.
    
    \item \textbf{Prompt refinement:} Appending new phrases compositionally integrates new acoustic elements without disrupting the primary acoustic scene.
\end{itemize}

As illustrated in Figure~\ref{fig:semantic_control}, these prompt-driven modifications confirm that our framework successfully retains the multi-step teacher's disentangled semantic representations, enabling highly controllable generation in a single forward pass.
}


{
\section{Limitations and Future Work}
\label{app:limitations}

\textbf{Limitations.}
SwiftAudio currently generates fixed-length audio segments of up to 10 seconds, following the duration setting of the underlying latent audio backbone and evaluation protocol. This limits its applicability to longer and more temporally structured audio scenes such as extended events or multi-stage acoustic narratives.

In addition, as a general text-to-audio generation model trained from caption-only supervision, SwiftAudio focuses on producing acoustically plausible sound events rather than linguistically controlled speech. For prompts such as “a man speaking,” the model typically produces realistic human voice-like sounds, but the spoken content may not correspond to a specific or identifiable language, since explicit phonetic or lexical supervision is not enforced.

\textbf{Future work.}
A promising direction is to extend SwiftAudio toward one-step semantic audio editing, enabling fast prompt-based modification of existing audio in a single forward pass. Combined with localized conditioning and attention control, this could support practical workflows such as sound replacement, attribute editing, and scene refinement with minimal computational cost.
}

\section{Conclusion}
This paper introduced \textit{SwiftAudio}, a one-step text-to-audio generation framework trained through audio-free distillation from a pretrained diffusion teacher. By adapting Variational Score Distillation to the audio domain and incorporating temporal smoothness regularization, the proposed method learns a caption-conditioned generator without paired text--audio supervision. Experimental results on AudioCaps and Clotho show that SwiftAudio achieves state-of-the-art performance among strict one-step methods while remaining competitive with multi-step diffusion systems. Furthermore, strong performance obtained from only approximately 45K captions highlights the effectiveness and data efficiency of the proposed distillation framework. 

\section*{Acknowledgment}
A generative AI tool, ChatGPT, was used only for limited language polishing, including grammar, style, and clarity improvements. It was not used to generate research ideas, methods, experimental results, analyses, or conclusions. All AI-assisted edits were reviewed and verified by the authors, who retain full responsibility for the manuscript content.

\bibliographystyle{IEEEtran}
\bibliography{mybib}

\section{Biography Section}

\begin{IEEEbiography}[{\includegraphics[width=1in,height=1.25in,clip,keepaspectratio]{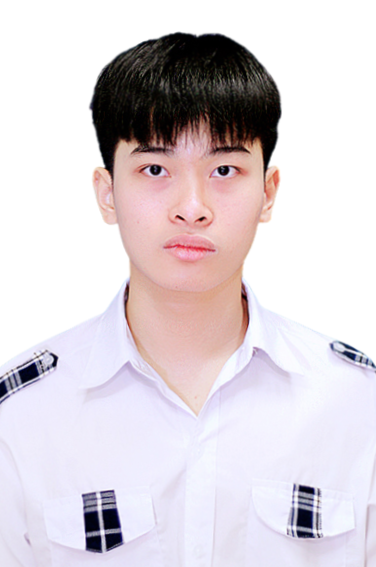}}]{Binh Mai}
received an Engineering degree with Honors in Information Technology from the Posts and Telecommunications Institute of Technology (PTIT), Hanoi, Vietnam. He is currently a Teaching Assistant at the Faculty of Artificial Intelligence, PTIT. His research interests include generative models, multimodal models, and audio processing.
\end{IEEEbiography}

\begin{IEEEbiography}[{\includegraphics[width=1in,height=1.25in,clip,keepaspectratio]{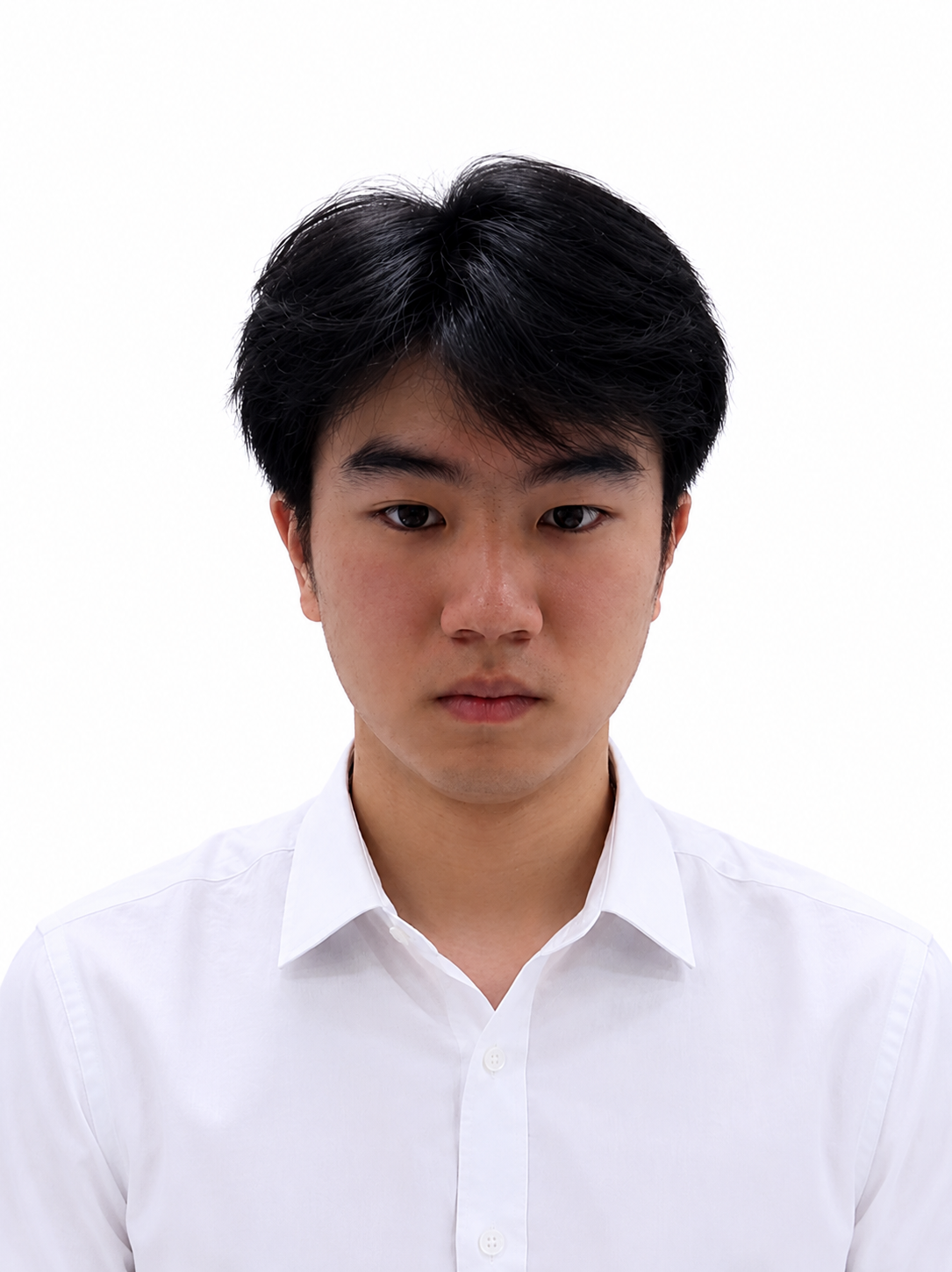}}]{Tran Quoc Bao Le}
is currently pursuing an Engineering degree at the Posts and Telecommunications Institute of Technology (PTIT), Hanoi, Vietnam. His research interests include deep learning, medical image reconstruction, generative modeling, and AI-powered applications. He is particularly interested in large language models, diffusion models, text-to-audio generation, and educational AI systems, with the goal of building efficient and accessible artificial intelligence solutions.
\end{IEEEbiography}
\begin{IEEEbiography}[{\includegraphics[width=1in,height=1.25in,clip,keepaspectratio]{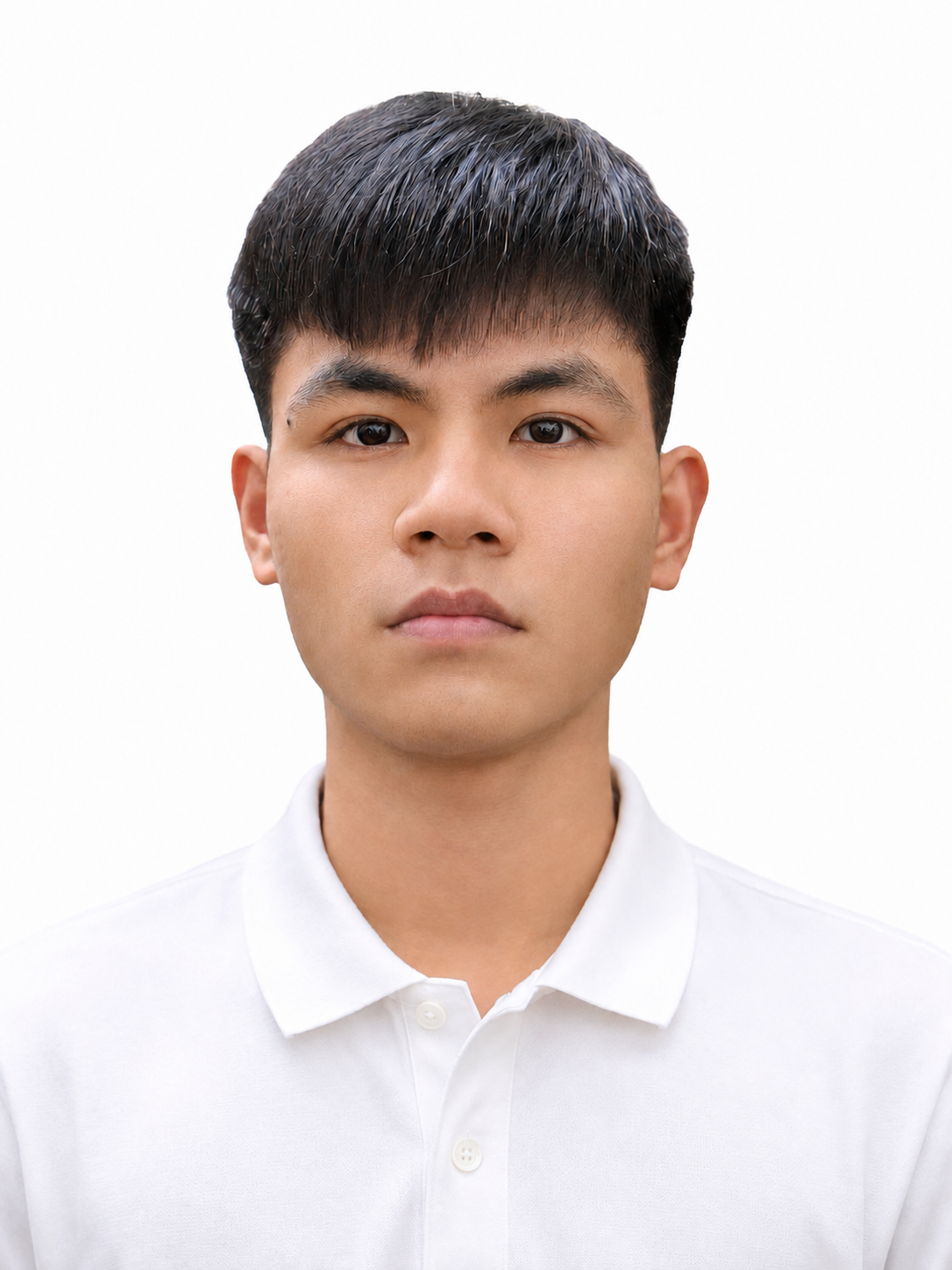}}]{Hung Dinh}
is currently pursuing an Engineering degree at the Posts and Telecommunications Institute of Technology (PTIT), Hanoi, Vietnam. His research interests include natural language processing, computer vision, generative modeling, and efficient deep learning. He is particularly interested in large language models, diffusion models, multimodal learning, and text-to-audio generation, with the goal of building scalable and data-efficient AI systems.

\end{IEEEbiography}

\begin{IEEEbiography}[{\includegraphics[width=1in,height=1.25in,clip,keepaspectratio]{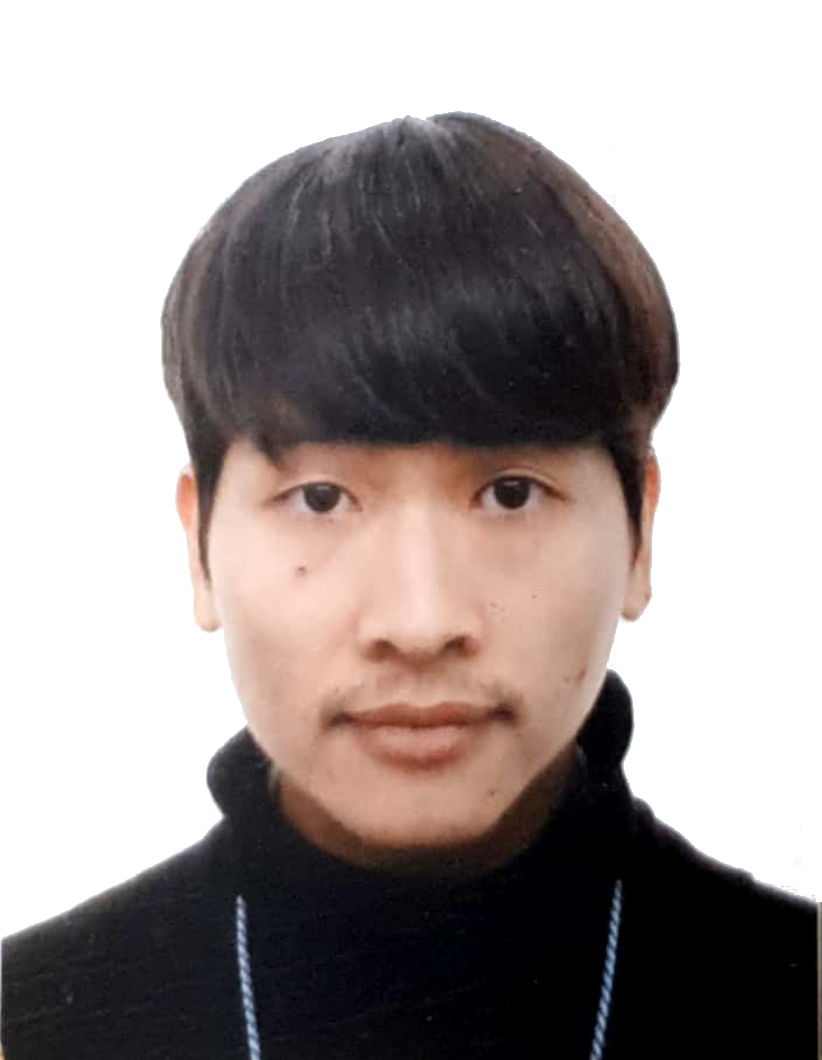}}]{Dr. Cong Tran}  received his doctoral degree in computer science from Dankook University, Yongin,
Republic of Korea, in 2021. He previously received his M.Sc. in computer science in 2014 and his B.Sc. in network and communication in 2009 from Vietnam National University, Hanoi, Vietnam. Since September 2021, he has been with the Faculty of Information Technology, Posts \& Telecommunication Institute of Technology, Hanoi, Vietnam, as a lecturer. His research interests include social network analysis, data mining, and machine learning.
\end{IEEEbiography}

\vfill

\appendices

\section{Evaluation checkpoints.}\label{app:checkpoints}
To ensure reproducibility and transparency, we provide the checkpoints used for evaluation in all comparative experiments.
All reported results in Table~1 of the main paper are obtained by directly evaluating these checkpoints.

The checkpoints are available at:
\begin{itemize}
    \item \textbf{Auffusion}: \url{https://huggingface.co/auffusion/auffusion-full-no-adapter}
    \item \textbf{AudioLCM}: \url{https://huggingface.co/liuhuadai/AudioLCM}
    \item \textbf{AudioLDM2}: \url{https://huggingface.co/cvssp/audioldm2}
    \item \textbf{ConsistencyTTA}: \url{https://huggingface.co/Bai-YT/ConsistencyTTA}
\end{itemize}

\section{Subjective Evaluation Details}\label{app:mos}

To ensure the transparency and reliability of our subjective evaluation, we provide a detailed description of the experimental protocol and the assessment interface used in this study.

\subsection{Participant Selection}
We recruited 31 participants for the Mean Opinion Score (MOS) tests. A critical requirement for selection was \textbf{full proficiency in English reading and comprehension}, ensuring that all evaluators could accurately interpret the nuances of the natural language captions provided in the AudioCaps dataset. This linguistic competence is essential for providing valid ratings for the Text Relevance (REL) metric. Participants were instructed to use high-quality headphones in a quiet environment to ensure optimal listening conditions.

\subsection{Evaluation Protocol}
The subjective test was conducted using a double-blind procedure to eliminate potential biases. Audio samples from SwiftAudio and the comparative baselines were presented in a randomized order, with all model identities hidden from the participants. 

For each sample, evaluators provided scores on a 5-point Likert scale (1: Bad to 5: Excellent) based on two criteria:
\begin{itemize}
    \item \textbf{Overall Quality (OVL):} Assessing the perceptual clarity, naturalness, and absence of unwanted artifacts in the audio.
    \item \textbf{Text Relevance (REL):} Assessing how well the generated audio aligns with the semantic content of the provided text prompt.
\end{itemize}

\subsection{Assessment Interface}
\begin{figure}[h]
    \centering
    \begin{minipage}{0.48\textwidth}
        \centering
        \includegraphics[width=\linewidth]{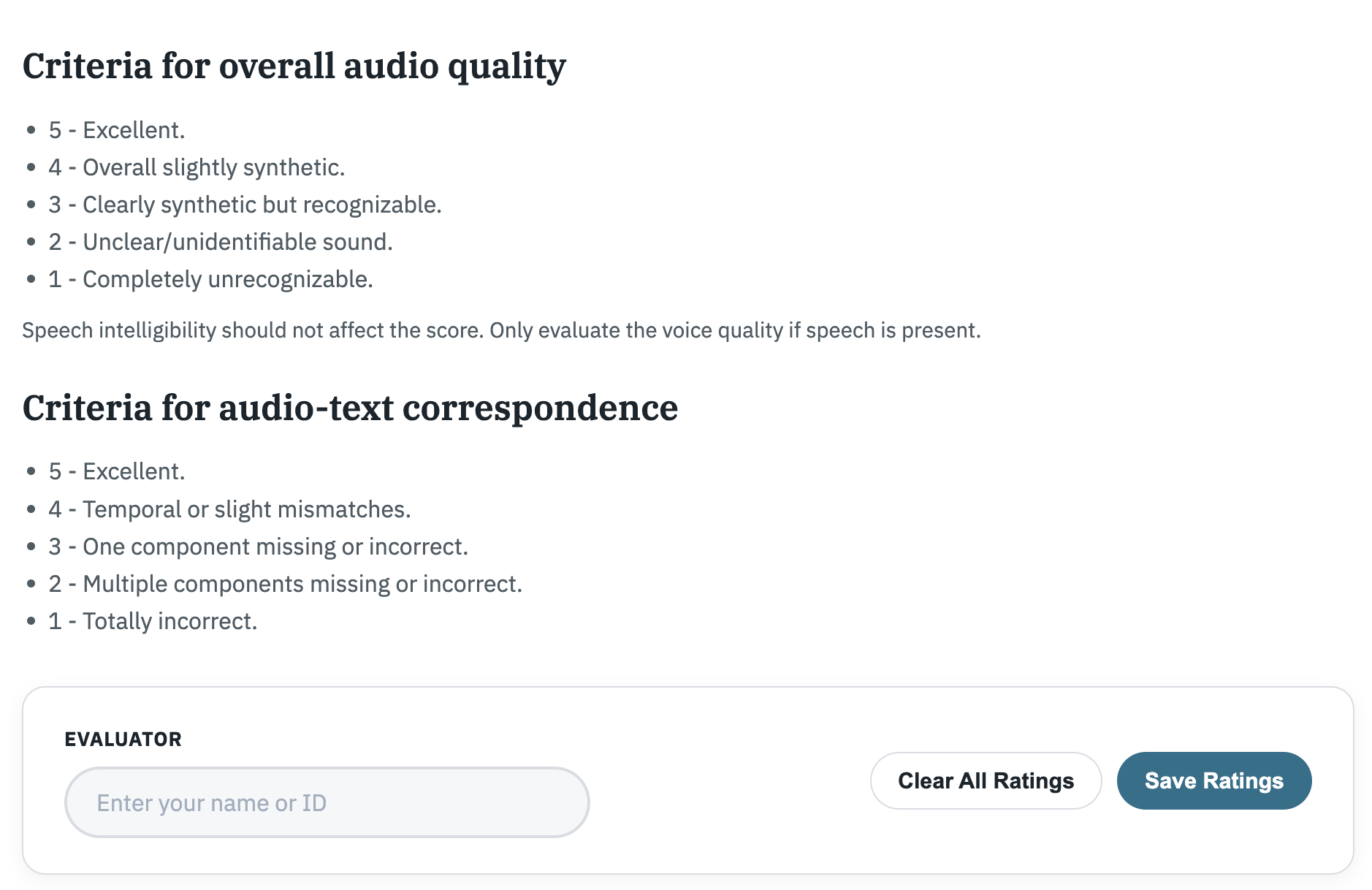}
        \caption{The assessment interface for rating audio quality and relevance based on the provided text prompt.}
        \label{fig:mos_interface_1}
    \end{minipage}
    \hfill
    \begin{minipage}{0.48\textwidth}
        \centering
        \includegraphics[width=\linewidth]{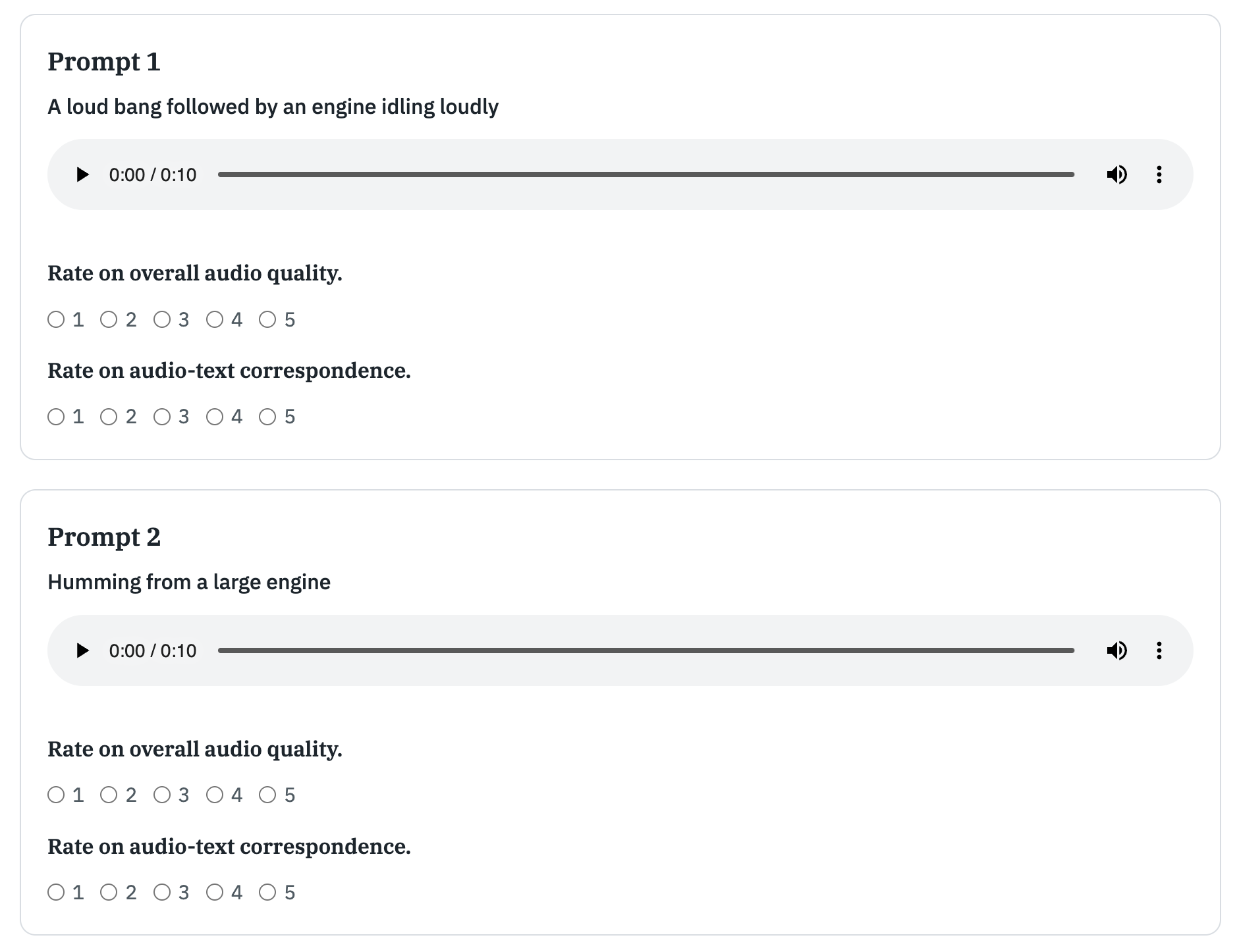}
        \caption{Instruction page for participants, defining the rating criteria and system requirements.}
        \label{fig:mos_interface_2}
    \end{minipage}
\end{figure}

The evaluation was performed via a dedicated web-based interface, as shown in Figures \ref{fig:mos_interface_1} and \ref{fig:mos_interface_2}. The interface provided clear instructions and allowed participants to listen to each sample multiple times before submitting their ratings. We monitored the time spent on each assessment to ensure that the evaluations were conducted with sufficient attention.

\section{Zero-shot Evaluation Protocol on Clotho}

The Clotho dataset \cite{drossos2020clotho} is used to evaluate the out-of-domain generalization ability of SwiftAudio. Unlike AudioCaps \cite{kim2019audiocaps}, Clotho recordings typically span 15--30 seconds, whereas all compared text-to-audio models generate fixed-length 10-second audio clips. To ensure a fair comparison, we randomly crop a 10-second segment from each reference audio sample prior to evaluation.

Each Clotho recording is accompanied by five human-written captions. Because the selected 10-second segment may only contain a subset of the acoustic events present in the full recording, not all captions necessarily remain equally representative after cropping. Directly selecting a caption at random would therefore introduce additional noise into the evaluation protocol.

To reduce caption-segment mismatch, we adopt a retrieval-based caption selection strategy using CLAP \cite{laionclap2023}. Given a cropped audio segment $a$ and its five associated captions ${c_i}_{i=1}^{5}$, we compute the CLAP similarity score between the audio embedding and each caption embedding:

\[
s_i = \mathrm{CLAP}(a, c_i).
\]

The caption with the highest similarity score is selected as the evaluation prompt:

\[
c^\star = \arg\max_{i \in {1,\dots,5}} s_i.
\]

The selected caption $c^\star$ is then provided as the text prompt to all evaluated generation models. The generated audio is compared against the corresponding cropped audio segment using the same objective metrics described in the main paper.

We emphasize that this procedure does not use any information from the generated audio and serves only to identify the caption that best matches the reference segment. The protocol reduces ambiguity caused by multiple valid descriptions and yields a more reliable estimate of text-audio alignment under zero-shot evaluation.

\section{Semantic Control}\label{app:p2p}

Inspired by prior work on prompt-based semantic editing in text-to-image generation \cite{hertz2023prompt2prompt},
we design a set of qualitative evaluations to probe semantic granularity and editability in SwiftAudio.

\subsection{Word Swapping}

\begin{figure*}[hbt!]
  \centering
  \includegraphics[width=\textwidth]{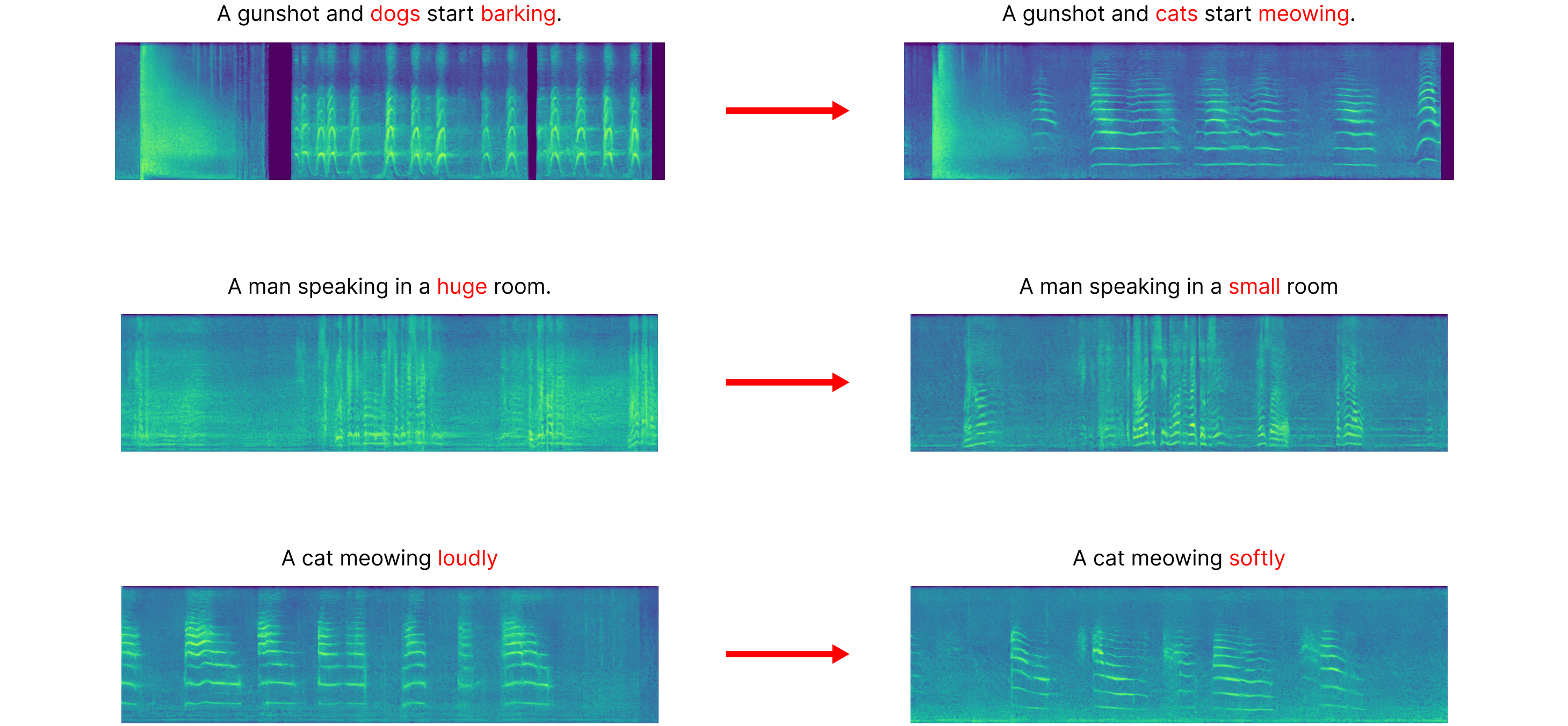} 
  \caption{Visualizing semantic control via word swapping. SwiftAudio precisely modifies specific sound events and acoustic properties (e.g., source, environment, dynamics) while maintaining overall scene consistency in a single inference step.}
  \label{fig:word_swap}
\end{figure*}

To evaluate the semantic granularity and controllability of SwiftAudio, we conduct a series of \emph{word swapping} experiments. As illustrated in Figure~\ref{fig:word_swap}, modifying specific keywords within the text prompt—e.g., transitioning from ``\red{dogs barking}'' to ``\red{cats meowing}''—results in a faithful and instantaneous transformation of the audio content.

A key observation is that SwiftAudio achieves these adjustments while preserving the global acoustic context. For instance, replacing the room size or sound intensity leaves the primary sound source intact, suggesting that the model has learned a disentangled latent representation. These modifications occur within a single generation step, bypassing the need for complex editing operations or iterative refinement typical of diffusion-based models.

Furthermore, the impact of acoustic modifiers (e.g., ``\red{loudly}'' vs. ``\red{softly}'') is clearly visible in the Mel-spectrograms, manifested as distinct shifts in energy distribution and spectral intensity. This level of fine-grained control via natural language demonstrates SwiftAudio's potential as a highly intuitive and efficient tool for one-step audio editing.

\subsection{Attention Reweighting for Intensity Control}

\begin{figure*}[hbt!]
  \centering
  \includegraphics[width=\textwidth]{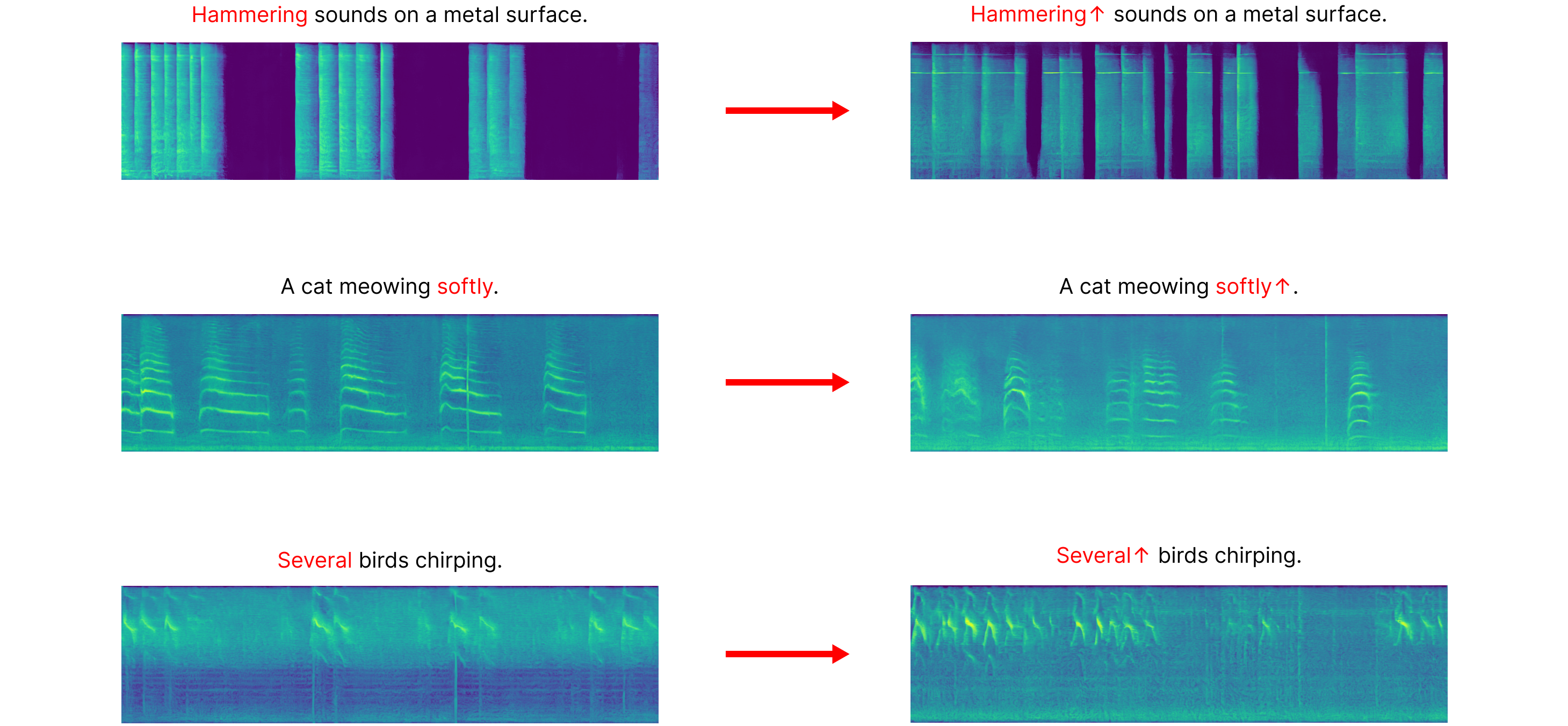} 
  \caption{Effect of attention reweighting. Increasing the weight of specific tokens enhances the intensity and presence of the target sound in the synthesized output.}
  \label{fig:reweight}
\end{figure*}

We further investigate the model's sensitivity to prompt emphasis by applying attention reweighting (denoted by \red{$\uparrow$}). Figure~\ref{fig:reweight} illustrates that increasing the attention weight of specific keywords leads to direct and intuitive changes in the synthesized audio: for "\red{Hammering}", the impacts become more frequent and more intense; for "\red{softly}", the sound becomes more delicate with shorter duration; and for "\red{Several}", the number of bird chirps increases significantly. These results demonstrate that the model accurately interprets attention weights to modulate sound intensity and density, proving that our VSD-based distillation effectively preserves the rich semantic understanding and fine-grained controllability of the original teacher model.

\subsection{Word Refinement}

Beyond simple keyword replacement, we further evaluate SwiftAudio under a more realistic and challenging setting: \emph{word refinement}. In this scenario, additional semantic phrases are incrementally appended to an existing prompt, preserving the core sound event while enriching the acoustic scene with auxiliary interactions or background sources.

\begin{figure*}[hbt!]
  \centering
  \includegraphics[width=\textwidth]{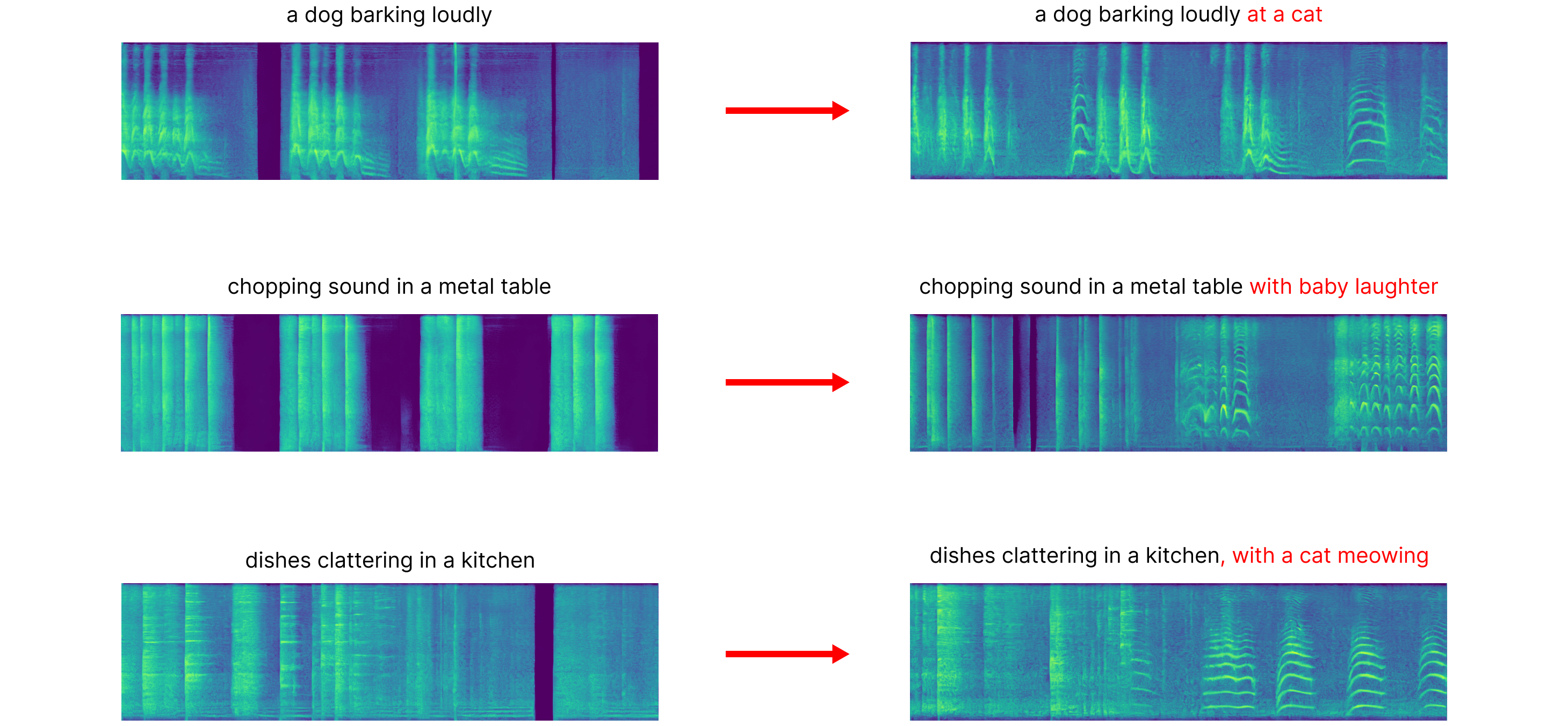} 
  \caption{Word refinement results in one-step text-to-audio generation. Starting from a base prompt (left), additional semantic phrases are appended (right), leading to localized and semantically consistent changes in the Mel-spectrograms. SwiftAudio preserves the primary sound event while compositionally integrating refined concepts such as secondary sound sources or interactions.}
  \label{fig:word_refine}
\end{figure*}

Figure~\ref{fig:word_refine} presents several representative examples. Starting from a base prompt such as ``a dog barking loudly,'' refining it to ``a dog barking loudly \red{at a cat}'' introduces new harmonic structures and temporal patterns associated with the secondary sound source, while the original barking characteristics remain clearly preserved. Similarly, augmenting ``chopping sound on a metal table'' with ``\red{with baby laughter}'' yields additional high-pitched, speech-like harmonic components that coexist with the sharp transient patterns of the chopping sound.

Importantly, these refinements do not disrupt the global temporal layout of the audio. The primary sound event remains dominant, while the refined semantic concepts are integrated in a compositional manner. This suggests that SwiftAudio performs localized semantic augmentation rather than regenerating the entire acoustic scene from scratch.

A more challenging case is shown in refining ``dishes clattering in a kitchen'' to ``dishes clattering in a kitchen, \red{with a cat meowing}.'' Here, SwiftAudio successfully injects distinct vocal-like harmonic contours into an otherwise noise-dominated spectrogram, demonstrating its ability to disentangle and recombine heterogeneous sound sources under a single-step generation regime.

Overall, these results highlight SwiftAudio’s capability for fine-grained and interpretable semantic refinement via natural language prompts, enabling intuitive control over complex auditory scenes without iterative editing or multi-stage synthesis.

\end{document}